\documentclass[preprint,12pt,authoryear]{elsarticle}
\usepackage{amssymb}
\usepackage{algorithm}
\usepackage{algpseudocode}
\usepackage{graphics}
 \usepackage{algorithm}
\usepackage[english]{babel}
\usepackage{indentfirst,graphicx}
\usepackage{amssymb,amsmath,latexsym,amsthm}
\usepackage{hyperref}
\usepackage{longtable}
\usepackage{multirow}
\usepackage{url}
\usepackage{color}
\def\T{{\footnotesize {^{_{\sf T}}}}} 

\newcommand{\bb}{\textcolor{blue}}
\journal{EcoSta journal}

\begin{document}
\begin{frontmatter}

 
\title{On approximate robust confidence distributions}

\author{Bortolato, E., Ventura, L. \\[0.5cm]
{\footnotesize Department of Statistical Sciences \\ University of Padova \\
Via C. Battisti 241, 35121 Padova, Italy} \\
{\footnotesize {\tt elena.bortolato.1@phd.unipd.it, ventura@stat.unipd.it}}}


\begin{abstract}
A confidence distribution is a complete tool for making frequentist inference for a parameter  of interest $\psi$ based on an assumed parametric model. Indeed, it allows to reach point estimates, to assess their precision, to set up tests along with measures of evidence for statements of the type "$\psi > \psi_0$" or "$\psi_1 \leq \psi \leq \psi_2$", to derive confidence intervals, comparing the parameter of interest with other parameters from other studies, etc.  

The aim of this contribution is to discuss robust confidence distributions derived from unbiased $M-$estimating functions, which provide robust inference for $\psi$  when the assumed distribution is just an approximate parametric model or in the presence of deviant values in the observed data.
Paralleling likelihood-based results and extending results available for robust scoring rules, we first  illustrate how robust confidence distributions can be derived from the asymptotic theory of robust pivotal quantities. Then, we discuss the derivation of robust confidence distributions via simulation methods. An application and a simulation study are illustrated in the context of non-inferiority testing, in which null hypotheses of the form $H_0: \psi \leq \psi_0$ are of interest.
 \end{abstract}


\end{frontmatter}


\section{Introduction} 

In recent years there has been considerable interest in frequentist inference based on confidence distributions (CDs) and confidence curves (CCs); see, among others, Xie and Singh (2013),  Schweder and Hjort (2016), Hjort and Schweder (2018), and references therein. In practice, a confidence distribution analysis is much more informative than providing a $(1-\alpha)$\% interval or a $p$-value for an associated hypothesis test. 

\begin{figure}
\begin{center}
\includegraphics[scale=0.7,height=7cm]{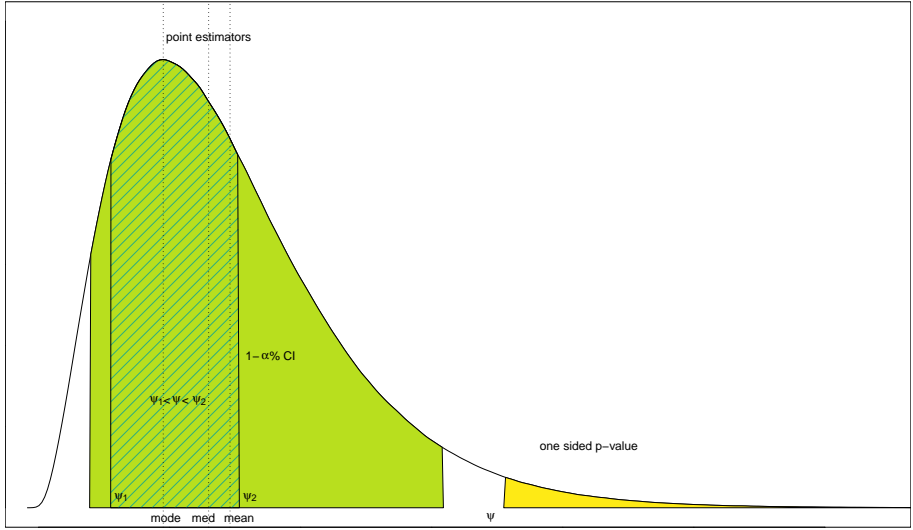}
\vspace{-0.2cm}
\caption{{\small Illustration of making inference on the scalar parameter of interest $\psi$ using a confidence density: point estimators (mode, median, mean), $(1-\alpha)\%$ quantile-type confidence intervals, one-sided $p$-value and measure of evidence for $\psi_1<\psi<\psi_2$.}}
\label{fig2}
\end{center}
\end{figure}

Let the scalar parameter of interest be  $\psi$. With inference on $\psi$ we shall understand statements of the type "$\psi > \psi_0$" or "$\psi_1 \leq \psi \leq \psi_2$", etc., where $\psi_0$, $\psi_1$  and
 $\psi_2$ are given fixed values. To each statement, a CD allows us to associate how much confidence the data have in the statement.
The plot in Fig. \ref{fig2} gives an illustration on making inference using a confidence density for a scalar parameter of interest $\psi$: point estimators, $(1-\alpha)$\% confidence interval, one-sided $p$-value and measure of evidence for "$\psi_1 \leq \psi \leq \psi_2$".

The standard theory for CDs evolves around the use of likelihood methods for a scalar parameter of interest $\psi$ of a parametric model. Typically, to first-order, CD inference may be based on familiar large-sample theory for the maximum likelihood estimator (MLE), the Wald statistic and the likelihood-ratio test.
However, it is well-known that likelihood-based methods are not robust when the assumed distribution is just an approximate parametric model or in the presence of deviant values in the observed data. In this case, it may be preferable to base inference on procedures that are more resistant, that is which specifically take into account the fact that the assumed models used by the analysts are only approximate. In order to produce statistical procedures that are stable with respect to small changes in the data or to small model departures, robust statistical methods can be considered (see, e.g., Hampel {\em et al.}, 1986, Huber and Ronchetti, 2009, Heritier {\em et al.}, 2009, and Farcomeni and Ventura, 2012, and references therein). 

The aim of this paper is to discuss the derivation of robust confidence distributions, together with their associated densities. Paralleling likelihood-based results, we first  illustrate that asymptotic robust CDs can be obtained by using pivotal quantities derived from unbiased $M$-estimating functions, extending the theory of robust scoring rules discussed in Hjort and Schweder (2018)   and Ruli {\em et al.} (2022) to the more general setting of $M$-estimating functions. Secondly, we  explore two simulation-based approaches to derive robust CDs. The first one relies  on a frequentist reinterpretation of Approximate Bayesian Computation (ABC) techniques (see, e.g.,  Rubio and Johansen, 2013, and Bee {\em et al.}, 2017, Ruli {\em et al.}, 2020, Thornton {\em et al.}, 2022). The second approach leverages a Montecarlo rejection algorithm for obtaining a one-sided $p$-value function. An application and a simulation study are illustrated in the context of non-inferiority testing (see, e.g., Rothman {\em et al.}, 2012), in which null hypotheses of the form $H_0: \psi \leq \psi_0$ are of interest in order to establish if a new product is not unacceptably worse than a product already in use. Finally, in conclusion, we mention the possibility to resort to a non-parametric derivation of CDs based on integral probability semimetrics (Muller, 1997) or pseudo-metrics (Huber and Ronchetti, 2009, Chap.\ 2). This approach makes use of discrepancies among cumulatives density functions as nonparametric pivots for performing inference on $\psi$, and an example is illustrated in a misspecification context, as in Legramanti {\em et al.} (2022).

The paper is organized as follows. Section 2  reviews  some background on CDs. Section 3 discusses the derivation of first-order robust CDs from unbiased $M$-estimating functions and of simulation-based CDs. An application and a simulation study are presented in Section 4 in the context of non-inferiority testing. Finally, Section 5 mentions the non-parametric derivation of CDs based on integral probability semimetrics, and Section 6 discusses some concluding remarks.


\section{
Background on confidence distributions}


\subsection{Approximate likelihood-based confidence distributions}

Consider a sample $y=(y_1,\ldots,y_n)$ of size $n$ from  a random variable $Y$ with assumed parametric model $f(y;\theta)$, indexed by a
 $d-$dimensional parameter $\theta$. Let $\theta=(\psi,\lambda)$, where $\psi$ is a scalar parameter of primary interest and $\lambda$ represents the remaining $(d-1)$  nuisance parameters.  
 
A recent definition of a confidence curve $cc(\psi)=cc(\psi,y)$ for $\psi$ can be found, among others, in Xie and Singh (2013) and Schweder and Hjort (2016); see also references therein. Let $\theta_0=(\psi_0,\lambda_0)$ the true parameter point. Then, the random variable $cc(\psi_0)=cc(\psi_0,Y)$ should have a uniform distribution on the unit interval and
\begin{eqnarray*}
P_{\theta_0} (cc(\psi_0,Y) \leq \alpha ) = \alpha, \quad  \text{for all} \, \, \alpha.
\end{eqnarray*}
Thus confidence intervals can be read off, at each desired level. When $\alpha$ tends to zero the confidence interval tends to a single point, say $\psi^*$, the zero-confidence level estimator of $\psi$. In regular cases, $cc(\psi)$ is decreasing to the left of $\psi^*$ and increasing to the right, in which case the confidence curve $cc(\psi)$ can be uniquely linked to a full confidence distribution $C(\psi)=C(\psi, y)$, via
\begin{eqnarray*}
cc(\psi) = |1-2C(\psi,y)| = \left\{ \begin{array}{ll}
1-2C(\psi,y), & \text{if}  \, \, \psi \leq \psi^* \\
2C(\psi,y)-1, & \text{if}  \, \, \psi \geq \psi^*.
\end{array} \right.
\end{eqnarray*}
With $C (\psi)$ a CD, $[C^{-1} (0.05), C^{-1} (0.95)]$ becomes an equi-tailed 90\% confidence interval. Also, solving $cc(\psi) = 0.90$ yields two cut-off points for $\psi$, precisely those of a 90\% confidence interval.  

A general recipe to derive a CD is based on pivotal quantities. Suppose $q(\psi;y)$ is a function monotone increasing in $\psi$, with a distribution not depending on the underlying parameter, i.e.  $q(\psi;y)$ is a pivot (Barndorff-Nielsen and Cox, 1994). Thus $Q(x) = P_\theta (q(\psi ;Y ) \leq x)$ does not depend on $\psi$, which implies that
\begin{eqnarray}
C(\psi) = Q(q(\psi;y))
\label{disq}
\end{eqnarray}
is a CD.  The corresponding confidence density for $\psi$ is
\begin{eqnarray*}
cd(\psi) =  \frac{\partial Q(q(\psi;y))}{\partial q(\psi;y)} \, \frac{\partial q(\psi;y)}{\partial \psi}.
\end{eqnarray*}
If the natural
pivot is decreasing in $\psi$, then $C(\psi) =1-Q(q(\psi;y))$. 

In the likelihood framework, there are well-working large-sample approximations for the behaviour of pivotal quantities and these lead to constructions of CDs. For instance,  if $\hat\psi$ is the MLE of $\psi$, then the CD is derived from the profile Wald statistic 
\begin{eqnarray}
w_p(\psi) = \frac{\hat\psi-\psi}{\sqrt{j_p(\hat\psi)^{-1}}},
\label{cdwald}
\end{eqnarray}
with $j_p(\psi)$ profile observed information,
and it coincides with the asymptotic first-order Bayesian posterior distribution for $\psi$ (see, for instance, Ruli and Ventura, 2021). 

A pivotal quantity that typically works better than (\ref{cdwald}) is the following. Let $\ell(\theta)$ be the log-likelihood function for $\theta$, and let $\ell_p (\psi) = \ell(\psi,\hat\lambda_\psi)$ be the profile log-likelihood for $\psi$, where $\hat\lambda_\psi$ is the MLE for $\lambda$ given $\psi$. The profile  log-likelihood ratio test $W_p(\psi) = 2 (\ell_p(\hat\psi) - \ell_p(\psi))$, under mild regularity conditions, has an asymptotic null $\chi^2_1$ distribution. Hence $ \Gamma_1(W
_p(\psi)) \, \, \dot\sim \, \, U(0,1)$, with $\Gamma_1(\cdot)$ denoting the $\chi^2_1$ distribution function, and 
\begin{eqnarray}
C(\psi) \, \dot{=} \, \Gamma_1(W_p(\psi))\label{asyCD_loglik}
\end{eqnarray}
is a first-order asymptotic CD, which can reflect asymmetry and also likelihood multimodality in the underlying distributions, unlike the simpler Wald-type confidence distribution. Similarly, the profile likelihood root
\begin{eqnarray*}
r_p (\psi) = \text{sign} (\hat\psi-\psi) \sqrt{2(\ell_p(\hat\psi)-\ell_p(\psi))}
\label{r}
\end{eqnarray*}
can be used to derive a first-order CD, since it has a first-order standard normal null distribution. 
Improved CD inference based on higher-order asymptotics (see, among others, Severini, 2000, Reid, 2003, Brazzale {\em et al.}, 2007, and references therein) is discussed, for instance, in Schweder and Hjort  (2016, Chap.\ 7); see also Ruli and Ventura (2021). One key formula is the modified profile likelihood root
\begin{eqnarray}
r_p^* (\psi) = r_p(\psi) + \frac{1}{r_p(\psi)} \log \frac{q_p(\psi)}{r_p(\psi)},
\label{rstar}
\end{eqnarray} 
which has a third-order standard normal null distribution. In (\ref{rstar}), the quantity $q_p(\psi)$ is a suitably defined correction term (see, e.g., Severini, 2000, Chapter 9). In practice, $r^*_p(\psi)$ is a higher-order pivotal quantity obtained as a refinement of the likelihood root $r_p(\psi)$, which allow us to obtain  an asymptotically third-order accurate CD, i.e.\ with error of order $O(n^{-3/2})$. 

\subsection{Simulation-based confidence distributions}

In the framework of CDs obtained from pivotal quantities, when the distribution of the pivot (\ref{disq}) is not known, approximate confidence distributions can be obtained by bootstrapping {(Efron, 1979)}. Indeed, this method provides an estimate of the sampling distribution of a statistic, and this empirical sampling distribution can be turned into an approximate confidence distribution in several ways (see, e.g., Schweder and Hjort, 2016, Chap.\ 7).
 
Bootstrap  procedures can be distinguished into two major categories: parametric and non parametric. With parametric bootstrap, the seeked distribution is taken to be that of the statistics given pseudo-data, generated by a consistent  estimate of the parametric model. With the second, instead, the pseudo-data generating scheme is a multinomial resampling of data points.
The error in bootstrap confidence intervals  is in most of the cases of order $O_p(n^{-1})$, becoming of order  $O_p(n^{-3/2})$ when correction procedures  can be implemented, and for $t-$bootstrap (see DiCiccio and Romano, 1988,  { DiCiccio and Efron}, 1996, Schweder and Hjort, 2016, Chap.\ 7).

For instance, consider  a monotone transformation $h(\psi)$ of a scalar parameter of interest, and let  $q(\psi,y)=(h( \psi)- h(\hat\psi))/\hat\tau$  an approximate studentized-pivot, where $\hat \psi$ is the MLE of $\psi$  and $\hat \tau$ is a suitable estimate of the pivot standard deviation. Let $Q(\cdot)$ be the distribution function of $q(\psi,y)$. Then, a confidence distribution  for the parameter of interest is   $C(h(\psi)) = Q \left((h(\psi)- h(\hat \psi))/\hat\tau \right)$, with appropriate confidence density $cd(h(\psi))=\partial C(h(\psi))/\partial \psi$. When $Q(\cdot)$ is unknown, it can be estimated via bootstrapping. Let $h(\psi^*)$ and $\hat\tau^*$ be the result of bootstrapping, then the $Q(\cdot)$ distribution can be estimated as $\hat Q$, via bootstrapped values of $q^* = q(\psi^*, y^*)=(h(\psi)- h(\hat \psi^*))/\hat\tau^*$. The approximate CD is then 
$$
C_{t-boot} (\psi) = \hat Q \left( \frac{h(\psi)- h(\hat \psi)}{\hat\tau} \right).
$$
This bootstrap method applies even when $q(\psi,y)$ is not a perfect pivot, but is especially successful when it is, because $q^*$ then has exactly the same distribution $Q(\cdot)$ as $q(\psi,y)$. Note that the method automatically takes care of bias and asymmetry in $Q(\cdot)$.


\section{Robust confidence distributions}

\subsection{Asymptotic derivation from $M$-estimating functions}

The class of $M$-estimators is broad and includes a variety of well-known estimators. For example  it includes the MLE, the maximum composite likelihood estimator (see e.g.\ Varin {\em et al.}, 2011), estimators based on proper scoring rules (see, e.g., Dawid {\em et al.}, 2016, and references therein), and classical robust estimators (see e.g.\ Huber and Ronchetti, 2009, and references therein).

Under broad regularity conditions, an $M$-estimator $\tilde\theta$ is the solution of the unbiased estimating equation 
$$
g(\theta) =  \sum_{i=1}^n g(y_i;\theta) = 0
$$ 
and it is asymptotically normal, with mean $\theta$ and covariance matrix \linebreak
$V (\theta) = K(\theta)^{-1} J(\theta) (K(\theta)^{-1})^{\T}$, 
where $K(\theta) = E_\theta (\partial g(\theta)/\partial \theta^{\T})$ and $J(\theta) = E_\theta(g(\theta) g(\theta)^{\T} )$ are the sensitivity and the variability matrices, respectively.  The matrix $V_g(\theta) = V(\theta)^{-1}$ is known as the Godambe information and its form is due to the failure of the information identity since, in general, $K(\theta) \neq J(\theta)$. Let us denote with $G(\theta)=\sum_{i=1}^n G(y_i;\theta)$ the function such that $g(\theta)$ is the gradient vector, i.e.\ $g(y;\theta)=\partial G(y;\theta)/\partial \theta$.

From the general theory of $M$-estimators, the influence function ($IF$) of the estimator $\tilde\theta$ is given by
\begin{eqnarray}
IF (y;\tilde\theta) = K(\theta)^{-1} g(y;\theta), 
\label{ifsco}
\end{eqnarray}
and it measures the effect on the estimator $\tilde\theta$ of an infinitesimal contamination at the point $y$, standardised by the mass of the contamination. The estimator $\tilde\theta$ is B-robust if and only if $g(y;\theta)$ is bounded in $y$. Note that the $IF$ of the MLE is proportional to the score function; therefore, in general, MLE has unbounded $IF$, i.e. it is not B-robust. 

%
%

Paralleling likelihood-based results, asymptotic robust inference on the scalar parameter of interest $\psi$ can be based on first-order pivots.  With the
partition $\theta=(\psi,\lambda)$, the $M$-estimating function is similarly
partitioned as $g(y;\theta)=(g_\psi(y;\theta),g_\lambda(y;\theta))$. Moreover,
consider the further partitions
$$
K = \left[
  \begin{array}{cc}
    K_{\psi \psi} & K_{\psi \lambda} \\
    K_{\lambda \psi} & K_{\lambda \lambda}
  \end{array}
\right], \quad K^{-1} = \left[
  \begin{array}{cc}
    K^{\psi \psi} & K^{\psi \lambda} \\
    K^{\lambda \psi} & K^{\lambda \lambda}
  \end{array}
\right],
$$
and similarly for $V_g$ and $V_g^{-1}$. Finally, let $\tilde\lambda_\psi$ be the constrained $M$-estimate of $\lambda$, let $\tilde\theta_\psi=(\psi,\tilde\lambda_\psi)$, and let $\tilde\psi$ be the $\psi$ component of $\tilde\theta$. Then, a profile Wald-type statistic for the $\psi$ may be defined as
$$
 w_{R}  (\psi) = (\tilde\psi - \psi) (\tilde{V}_g^{\psi \psi})^{-1/2},
$$
and it has an asymptotic $N(0,1)$ null distribution. Similarly, the profile score-type statistic
$$
w_{sR}  (\psi) = g_\psi (\tilde\theta_{\psi})^T K^{\psi \psi}
(V_g^{\psi \psi})^{-1} K^{\psi \psi} g_\psi (\tilde\theta_{\psi})
$$ 
has an asymptotic $\chi^2_1$ null distribution, while the asymptotic distribution of the profile  ratio-type statistic for $\psi$, given by $W_{R} (\psi) = 2 \left( G(\tilde\theta_{\psi}) - G(\tilde\theta) \right)$,
is $ \nu \chi^2_1$, where $\nu = (\tilde{K}^{\psi \psi})^{-1} \tilde{V}_g^{\psi \psi}$.  In view of this, for the adjusted profile ratio-type statistic to first-order it holds
  \begin{eqnarray*}
   W_{R}^{adj} (\psi) = \frac{W_{R} (\psi)}{\nu}\ \, \dot{\sim} \,  \chi^2_1.
    \label{swadj}
  \end{eqnarray*}
Finally, the adjusted profile root, analogous to (\ref{r}), can be defined as
\begin{eqnarray*}
r_{R} (\psi) = \text{sign} (\tilde\psi - \psi) \sqrt{ W_{R}^{adj} (\psi) },
\label{rsradj}
\end{eqnarray*} 
which has an asymptotic standard normal distribution. For the general theory of robust tests see Heritier and Ronchetti (1994). 

Paralleling results in Section 2.1 for likelihood based CDs, a recipe to derive an asymptotic CD from robust $M$-estimating functions is based on pivotal quantites, extending the theory illustrated for robust scoring rules in Hjort and Schweder (2018) and Ruli {\em et al.} (2022). To this end, let us denote with $q_R (\psi;y)$ a robust pivotal quantity, such as the profile Wald-type statistic $w_{R}  (\psi)$ or the adjusted profile scoring rule root $r_{R} (\psi)$. Then, 
\begin{eqnarray}
C_{R}^w(\psi) \, \dot{=} \, \Phi\left( (\psi - \tilde\psi) (\tilde{V}_g^{\psi \psi})^{-1/2} \right)
\label{cd1}
\end{eqnarray}
and
\begin{eqnarray}
C_{R}^r(\psi) \, \dot{=} \, \Phi \left( \text{sign} (\psi - \tilde\psi) \sqrt{ W_{R}^{adj} (\psi) } \right)
\label{cd2}
\end{eqnarray}
are first-order asymptotic CDs, and the corresponding confidence densities are, respectively,
$$
cd_{R}^w(\psi) \, \dot{=} \, \frac{\phi \left((\psi - \tilde\psi) (\tilde{V}_g^{\psi \psi})^{-1/2} \right)}{\sqrt{\tilde{V}_g^{\psi \psi}}}
$$
and
$$
cd_{R}^r(\psi)  \, \dot{=} \,  \phi \left( \text{sign} (\psi - \tilde\psi) \sqrt{ W_{R}^{adj} (\psi) } \right) \,  \left| \frac{\partial W_{R}^{adj} (\psi)^{1/2}}{\partial \psi} \right|,
$$
where $\phi(\cdot)$ is the density function of the standard normal distribution. Note that the Wald-type based confidence density $cd_{R}^w(\psi) $ coincides with the asymptotic first-order robust Bayesian posterior distribution for $\psi$ (see, e.g., Greco {\em et al.}, 2008, and Ventura and Racugno, 2016).

In practice, using for instance (\ref{cd2}), the confidence median is $\tilde\psi$ and an $(1-\alpha)$ equi-tailed confidence interval can be obtained as $\{\psi : |r_{R} (\psi)| \leq z_{1-\alpha/2} \}$, where $z_{1-\alpha/2} \}$ is the $(1-\alpha/2)$-quantile of the standard normal density. When testing, for instance, $H_0: \psi = \psi_0$ against $H_1: \psi < \psi_0$, the $p$-value is $p=C_{R}^r(\psi_0)$, while when testing
 $H_0: \psi = \psi_0$ against $H_1: \psi \neq \psi_0$ the $p$-value is $p = 2(1-\Phi(|r_{R}(\psi_0)|))$. Furthermore, a measure of evidence for a statement of the form "$\psi_1<\psi<\psi_2$" can be computed as $C_{R}^r(\psi_2)-C_{R}^r(\psi_1)$.
 
To study the stability of robust CDs, let us write the robust pivotal quantity more generally as $q_R (\psi; T(\hat{F}_n))$, where $\hat{F}_n$ is the empirical distribution function and $T(F)$ is the functional defined by the unbiased $M$-estimating equation $\int g(y;T(F)) \, dF(y)=0$, where $F=F(y;\theta)$ is the assumed parametric model. 
In CD  inference  the tail area, given by $C_R(\psi)  = \Phi (q_R (\psi; T(\hat{F}_n))$, plays a central role and thus  we can consider the tail area influence function (see, e.g., Field and Ronchetti, 1990, and Ronchetti and Ventura, 2001), given by
\begin{eqnarray}
TAIF(y;T) = \left. \frac{\partial}{\partial \varepsilon} \Phi (q_R (\psi; T(F_\varepsilon))) \right|_{\varepsilon=0},
\label{taif}
\end{eqnarray}
where $F_\varepsilon = (1-\varepsilon) F + \epsilon \Delta_y$ and $\Delta_y$ is the probability measure which puts mass 1 at the point $y$. The $TAIF(y;T)$ thus describes the normalized influence on the CD tail area of an infinitesimal observation at $y$ and,  by considering its supremum, it can be used to evaluate the maximum bias of the tail area on the $\varepsilon$-neighborhood of $F$. It can be shown that
\begin{eqnarray}
TAIF(y;T) & = &  \phi(q_R(\psi;T(F))) \, \, \frac{\partial q_R(\psi;T(F))}{\partial T(F)} \, \, \left. \frac{\partial T(F_\varepsilon)}{\partial \varepsilon} \right|_{\varepsilon=0},
\label{taif2}
\end{eqnarray}
where the last term in (\ref{taif2}) is the IF (\ref{ifsco}) of the $M$-estimator. Thus, the tail area influence function for the CD tail area at the statistical model $F$ is proportional to the $M$-estimating function and this gives an immediate  handle on robustness. Furthermore, it  is bounded with respect to $y$ when the $M$-estimating function is bounded.

The application of (\ref{cd1}) and (\ref{cd2}) in the particular context of a robust scoring rule has been discussed in Ruli {\em et al.} (2022). In particular, the Tsallis score (Tsallis, 1988) is considered, which is given by
\begin{equation*}
  \label{eq:tsallisscore}
  G(y;\theta) = (\gamma - 1) \int\!  f(y;\theta)^\gamma \, d y - \gamma f(y;\theta)^{\gamma-1},
  \quad \gamma>1,
\end{equation*}
with corresponding unbiased $M$-estimating function $g(\theta)=\partial G(y;\theta)/\partial \theta$ (Ghosh and Basu, 2013, Dawid {\em et al.}, 2016), and with the parameter $\gamma$ which gives a trade-off between efficiency and robustness. 

%


\subsection{Derivation of robust confidence distributions via simulation}
 
In the context of robust procedures, suitable modifications of the bootstrap have been explored for improving the stability of inference. In fact, both families of  parametric and non parametric bootstrap presents some drawbacks. For instance, with the non parametric bootstrap, despite  the direct specification of the data generating process is avoided and this robustifies the analysis, the distribution of estimators might be unstable since the amount of outliers
 of pseudo-samples can be higher than that in the original dataset. 
For remedying this drawback, some authors proposed the use of weigths associated to observations  before performing repeatedly likelihood maximization  under the assumed model, as in  weighted Likelihood Bootstrap (Newton and Raftery, 1994). Lyddon {\em et al.}  (2019) showed that the corresponding estimator is asymptotically normal and with covariance matrix with the classical structure of sandwich form. Also, Chen and Zhou (2020) introduced a  similar approach based on estimating equations and provided non-asymptotic guarantees for the resulting errors.
Moreover, a further problem when using a bootstrap approach together  with robust procedures as  $M$-estimating equations is the need of repeatedly solving numerically non-convex or complex optimization problems, which may be computationally expensive.

In this section we inspect two alternative methods for computing  CDs based on robust $M$-estimating functions, that go beyond some limitations of the bootstrap.
Broadly speaking, the first method is based on a frequentist reinterpretation of the ABC machinery (see, e.g., Bee {\em et al.}, 2017, Ruli {\em et al.}, 2020, Thornton {\em et al.}, 2022), whose properties have been derived by Rubio and Johansen (2013) in a general setup. The idea consists in generating candidate parameter values from an uniform distribution, computing a robust suitable summary statistic using the simulated data and then accepting only the parameter values such that the corresponding summary statistic is "close" to its observed counterpart (see Algorithm \ref{algo1}).

\begin{algorithm}[H]
 \caption{Accept-reject robust ABC}
 Input: proposal $p(\psi, \lambda)$, number of iterations $R$, robust summary statistic $t(\cdot)$, $t^{obs}=t(y^{obs})$, where $y^{obs}$ is the observed sample, tolerance $\varepsilon$, distance $\rho(\cdot;\cdot)$
  \begin{algorithmic}
  \For{$j \in 1,\ldots,R$}
       \State Sample $(\psi_j^*, \lambda_j^*) \sim p(\psi, \lambda)$
 and  $y_j^*\sim f(y;\psi_j^*, \lambda_j^*)$
 \State Compute $t_j^*=t(y_j^*)$ 
\State  Accept $\psi_j^*$ if $ \rho(t_j^*; t^{obs})\leq \varepsilon$ else reject 
 \EndFor
 \State resample  the accepted $(\psi^{*}, \lambda^*)$ with probability $\propto1/p(\psi^*,\lambda^*)$ \\
 \Return    robust approximate normalized pseudo-likelihood $\propto$ confidence density $\hat{cd}_R^{abc} (\psi)$  
 \end{algorithmic} \label{algo1} 
 \end{algorithm}
 
In Algorithm \ref{algo1}, the summary statistics of Soubeyrand and Haon-Lasportes (2015) or of Ruli {\em et al.} (2016, 2020) can be used. In particular, the first one is based directly on the $M$-estimator $\tilde\psi$ as the summary statistic $t(y)$ and a, possibly rescaled, distance among the observed and the simulated value of the statistic. In the second one, a rescaled version of the $M$-estimating function $g(\theta)$, evaluated at a fixed value of the parameter, is used as a summary statistic $t(y)$; this avoids repeated evaluations of the consistency correction involved in the $M$-estimating function. 
Note also that when using the $M$-estimator as a summary statistic, the algorithm for solving the estimating equation might not converge after a prefixed number of iterations, thus causing additional noise in the results.
%
%
For a single parameter of interest, with the partition $\theta=(\psi,\lambda)$, we propose to modify the algorithm of Ruli {\em et al.} (2020) by  using  the profile estimating equation $g_\psi (y;\psi,\lambda)$ and plugging in the value of proposals $\lambda^*$ for nuisance parameters used to generate pseudo-data.
The treatment of the nuisance parameters resembles as a generalized profile likelihood computation.
\bb{}

Note that, assuming the regularity assumptions of Soubeyrand and Haon-Lasportes (2015) and the usual regularity conditions on $M$-estimators (Huber and Ronchetti, 2009, Chap. 4), then for $n \to \infty$ the robust confidence densities derived via simulation are asymptotically equivalent to the Wald-type confidence density $cd_R^w(\psi)$. Moreover, following Ruli {\em et al.} (2020), if $g(y;\theta)$ is bounded in $y$, i.e. if the $M$-estimator is B-robust, then asymptotically the posterior mode, as well as other posterior summaries of the robust confidence density $\hat{cd}_R^{abc} (\psi)$ have bounded $IF$. 

The second method  for computing  CDs based on robust $M$-estimating functions  is similar but aims at computing directly a Montecarlo $p$-value  or a significance function using a different acceptance rule (see Bortolato and Ventura, 2022): again the parameter values are sampled from a uniform distribution, a robust summary statistic $t(y^*)$  using the simulated data is obtained, and then the proposed parameter is accepted if the robust summary statistic $t^*$ is greater of its observed counterpart $t^{obs}$ (see Algorithm \ref{algo2}).
For obtaining the confidence density, if the obtained CD is indeed monotone increasing, hence it is a proper CD, Algorithm \ref{algo3} can be used.

\begin{algorithm}[H]
 \caption{Accept-reject confidence curve and confidence distribution computing}
 Input: proposal $p(\psi, \lambda)$, number of iterations $R$, robust summary statistic $t(\cdot)$, $t^{obs}=t(y^{obs})$, where $y^{obs}$ is the observed sample.
  \begin{algorithmic}
  \For{$j \in 1,\ldots,R$}
       \State Sample $(\psi_j^*, \lambda_j^*) \sim p(\psi, \lambda)$ 
 and  $y_j^*\sim f(y;\psi_j^*, \lambda_j^*)$
 \State Compute $t_j^*=t(y_j^*)$ 
\State  Accept $\psi_j^*$ if $t_j^*\geq t^{obs}$ else reject 
 \EndFor
 \Return  $\psi^{*}\text{ with density }\propto cc_R(\psi)$ a confidence curve/distribution
 \end{algorithmic} \label{algo2} 
 \end{algorithm}

 \begin{algorithm}[H]
  \caption{Confidence density} 
  Input: robust confidence distribution $C_R(\psi),$ desired  size  $R$,
 grid of values  $G= \{ \frac{1}{R}, \frac{2}{R},\ldots ,\frac{R-1}{R}, 1\}$
 \begin{algorithmic}   
  \For{$j \in 1,\ldots,R$}
 \State  Compute the empirical quantiles of the CD: 
   $\psi_j^* = C_R^{-1}(G_j)$ 
   \EndFor \\
  \Return $\psi^*\sim \hat{cd}_R(\psi)$.  
 \end{algorithmic} \label{algo3}
 \end{algorithm}

As a final remark, we observe that, for obtaining stable results with the accept-reject schemes, it is suggestable to increase the number of proposals as the dimesion of the parameter space increases, in order to have resonable acceptance rate with ABC-type algorithm, and in general for obtaining more precise estimation of the confidence distributions.



\section{Applications to non-inferiority  tests}

The aim of this section is to introduce and apply CDs inference in the context of non-inferiority testing, in which interest is in establishing if a new product is not unacceptably worse than a product already in use. Applications of non-inferiority testing has revealed an attractive problem in  medical statistics, biostatistics, statistical quality control and engineering statistics, among others. Here we focus in non-inferiority clinical trials where the aim is to show that an experimental treatment is not (much) worse than a standard treatment. Clinical practice, however, is not the only field of application of these tests:  in comparing the performance of sensors in industrial environment, for instance, the  margin may be linked to some difference in costs due to  sensor functioning.
Other applications can be found in machine learning literature, where instead the meaningful
margin is related to the accuracy or to the speed in classification tasks.


In the process of evaluating the efficacy of an experimental treatment, it is common to develop studies in which the two arms are the new  and the standard therapy, respectively, rather than the new and the placebo. This is  because it is considered  unethical to deprive patients from a therapy that has already been proven to be beneficial. The underlying research hypothesis to be verified is that new therapies have equivalent or non-inferior efficacies to the ones currently in use. Both non-inferiority and superiority tests are examples of directional (one-sided) tests (see, e.g., D'Agostino {\em et al.}, 2003, Rothmann {\em et al.}, 2012, and references therein). In particular, the {\em non-inferiority test} wants to test that the treatment mean $\mu_N$ is not worse than the reference mean $\mu_S$ by more than a given equivalence margin $\delta$. The actual direction of the hypothesis depends on the response variable being studied. This question can be formulated into a test procedure for which the null hypothesis is 
$$
H_0: \mu_S - \mu_N \geq \delta,  {}
$$
where $\delta>0$ is the equivalence margin, when higher values of the response variable mean better results, versus
$$
H_1: \mu_S- \mu_N <\delta.
$$
The scalar parameter of interest in this context is thus $\psi= \mu_S - \mu_N$, and non-inferiority is claimed when the null hypothesis is rejected.

\begin{figure}
\centering
\includegraphics[]{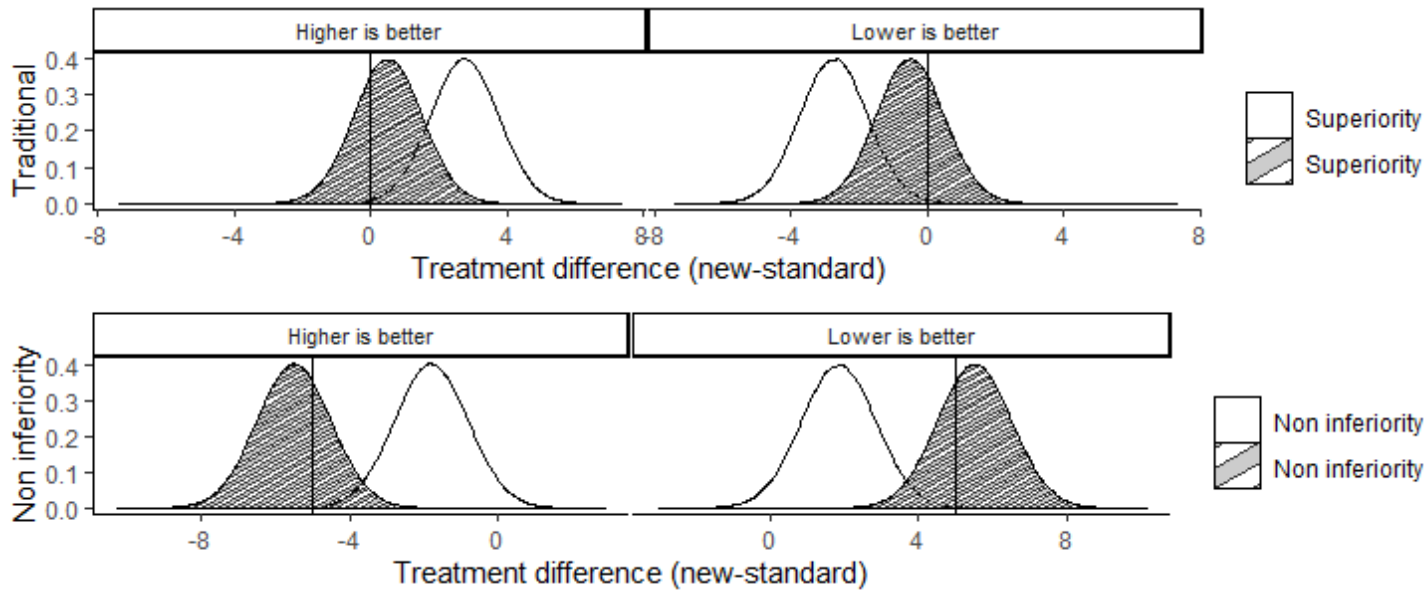}
\caption{Testing procedure with traditional comparative studies and non-inferiority studies using confidence densities: vertical lines represent the equivalence/non-inferiority margin ($\delta=-5$).} \label{tost}
\end{figure}

The equivalence margin $\delta$ corresponds to the practical acceptable difference and  should be pre-specified before the data is recorded (see e.g. Garret, 2003). An overly conservative margin might result in a high risk of not being able to claim non-inferiority when it actually is non-inferior. Conversely, overly liberal margins could result in a high risk of claiming non-inferiority when it actually is not non-inferior. A reasonable margin would be best derived from a combination of factors: the expected event rate, the duration of follow-up, and the number and nature of the events. However, arbitrary clinical judgment and the sponsor budget are of a great influence, resulting in a somewhat subjective non-inferiority margin. It is not clear in some situations how to perform the choice, and multiple thresholds could be plausible; in this respect, CDs are particularly useful to perform sensitivity analyses.  Indeed, in this situation a confidence distribution on the difference $\psi = \mu_S - \mu_N$ will simultaneously show the evidence of the $p$-value against the null for a series of values $\delta$, and decide for a reasonable  $\delta$ with the nominal control of the rejection level and possible alternatives. 

\begin{figure}[h]
\center
\includegraphics[scale=0.65]{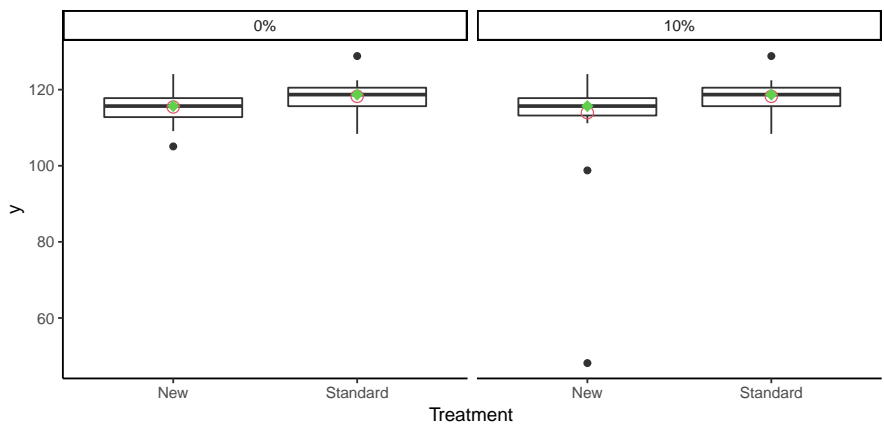}
\caption{Boxplots of the new treatment group and of the standard group under the two scenarios. Red dots indicate group means, green dots group medians.}
\label{boxplot}
\end{figure}

Here we consider  an example of trial where higher levels of the response variable mean that the new treatment is effective. The aim is verifying that the new treatment ($N$) is not unacceptably worse to the standard ($S$). Let us assume that  $n=80$ patients are randomized into two groups, and the model for the data is assumed to be 
\begin{equation}
Y_S= \mu_N + \psi+ u,  \\
\quad Y_N= \mu_N   +u, \quad u \sim N(0,\,\sigma^2). \label{model_trial}
\end{equation}
The normal distribution on the error term is often the basis of statistical analyses in medicine, genetics and in related sciences. Under this assumption, parametric inferential procedures based on the sample means, standard deviations, two-samples $t$-test, and so on, are the most efficient. However, it is well known that they are not robust when the normal distribution is just an approximate parametric model or in the presence of deviant values in the observed data (see, e.g., Farcomeni and Ventura, 2012).
In the framework described by (\ref{model_trial}), we inspect the effect of adding some contamination in the data of the new treatment group. In particular, in the contaminated scenario, $10\%$  of the error terms in the new treatment group are half-Cauchy distributed (see Fig. \ref{boxplot}).

\begin{figure}[h]
\center
\includegraphics[scale=0.625]{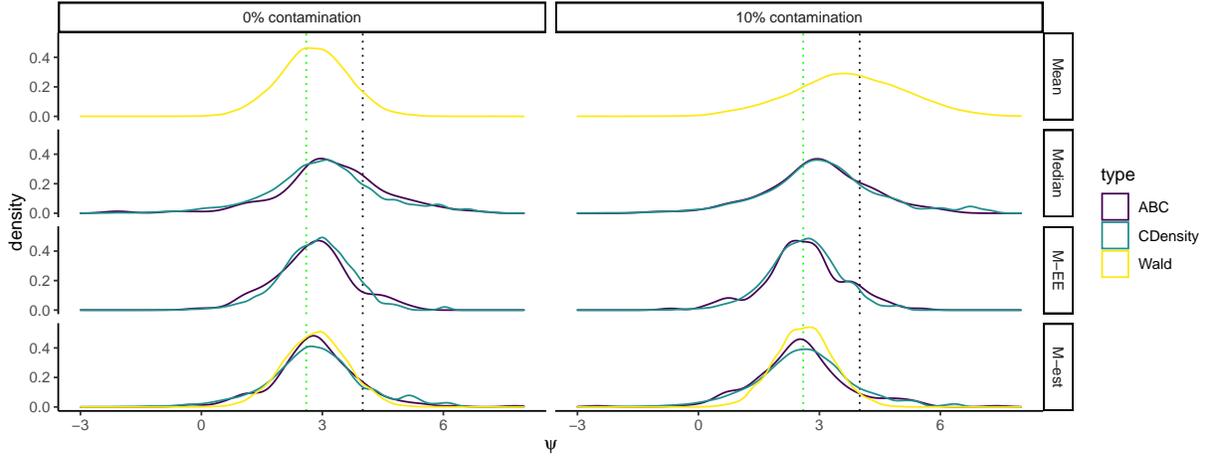}
\caption{Confidence densities for $\psi$ based on $10^5$ proposals,  without (left column) and with contamination (right column). Results according to a different types of statistic are reported for each row, inferential techniques are represented with different colors. Black vertical dotted line is $\delta$, green vertical dotted line is $\psi_0$.}
\label{varieCD}
\end{figure}

It is of interest to compare CDs inference for $\psi$ based on the following approaches (abbreviations are also used in Fig.\ \ref{varieCD} and in the following) used to derive confidence densities:
\begin{enumerate}
\item 
exact classical Wald-type confidence density based on $w_p(\psi)$, which is related to the classical two sample $t$-test (Wald/Mean) 
\item 
robust asymptotic Wald-type confidence density $cd_R^w(\psi)$ based on the Huber's estimator (Wald/M-test)
\item 
approximate confidence density based on ABC (Algorithm \ref{algo1}) with robust Huber's estimator as summary statistics (ABC/M-est)
\item 
approximate confidence density based on ABC (Algorithm \ref{algo1}) with the robust Huber's estimating equation as summary statistic (ABC/M-EE)
\item 
simulated confidence density (Algorithm \ref{algo2}) based on the robust Huber's estimator (CDensity/M-est)
\item 
simulated confidence density (Algorithm \ref{algo2}) based on  the robust Huber's estimating equation  (CDensity/M-EE)
\item 
approximate confidence density based on ABC (Algorithm \ref{algo1}) with the difference of medians  as summary statistics (ABC/Median)
\item 
simulated  confidence density (Algorithm \ref{algo2}) based on the difference of medians  (CDensity/Median)
\item
parametric bootstrap  confidence density (Boot/Basic)
\item
parametric bootstrap with normal intervals confidence density (Boot/Norm)
\item
parametric bootstrap with percentiles confidence  density (Boot/Perc)
\end{enumerate}
The nominal value of the mean difference between the treatment effects is $\psi_0$ is fixed to 2.6, and for simulation-based confidence distribution as well as for those obtained by the ABC-type algorithm we used $10^5$ proposals and a tolerance level of $0.1$. In the Huber's estimator we fix the tuning constant which controls the desired degree of robustness to 1.345, which imply that the estimator is 5\% less efficient than the corresponding MLE under the assumed model.

From the resulting confidence densities illustrated in Fig.\ \ref{varieCD} we note that, when the data come from the central model (left column) all the confidence densities are in reasonable agreement, even if the confidence densities based on the medians behave slightly worse, with a greater variability.  When the data are contaminated (right column), the non robust confidence density Wald/Mean is less trustworthy as it drifts away from the true parameter value (green dotted line). This is not the case however for the robust confidence densities which remains centred around the true parameter value. We further note that in the contaminated case, the robust confidence densities based on the $M$-estimating equation (ABC/M-EE and CDensity/M-EE) are the ones with the smallest variability.  For all these confidence densities, Table \ref{evidence} gives the measures of evidence for the equivalence margin $\delta$ taken equal to 4 (black dotted line in Fig. \ref{varieCD}), that is for the statement "$\psi>\delta$". As a reference, consider the result  derived by the exact $t-$distribution of the exact classical Wald-type confidence density in the non contaminated case, which  is $0.08$. The results, without and with the contamination, confirm the behaviour of the confidence densities in Fig.\ \ref{varieCD}, in particular the non robustness of the likelihood-based confidence density (Wald/Mean). The most stable values under contamination seem to be those obtained with M-EE approaches (0.09 with ABC/M-EE and 0.05 with CD/M-EE).  The same analysis could be done in principle, given the CD, for any margin $\delta$.

\begin{table}[h]
\centering
\begin{tabular}{|l|l|l|}
  \hline 
  Method &  $0\%$ cont. &  $10\%$ cont.   \\ 
  \hline 
  Wald/Mean & 0.08& 0.42 \\ 
Wald/M-test & 0.07& 0.03\\ 
ABC/Median & 0.24 & 0.20  \\ 
   ABC/M-EE  & 0.11 & 0.09 \\ 
   ABC/M-est & 0.12 & {0.09} \\   
  CDensity/Median & 0.20 &   0.21 \\ 
   CDensity/M-EE & 0.08&  0.05\\ 
   CDensity/M-est & 0.14 &   0.11 \\   
   \hline
\end{tabular}
\caption{Confidence measures of evidence for the null hypothesis $H_0: \psi>\delta$, with $\delta=4$ associated to Fig.\ \ref{varieCD}, without and with contamination.}\label{evidence}
\end{table}

\subsection{Simulation study}

For investigating the behaviour of the several confidence densities, we perform a simulation study under  two  sample sizes $n=40, 80$ (20, 40 per group) and  for each of them we investigate two scenarios: one in which the assumptions of the model in (\ref{model_trial}) are met by the true data generating model,
and the second one where $10\%$  of the error terms in the new treatment group are half-Cauchy distributed (as in Fig.\ \ref{boxplot}).
The families of methods to derive the confidence densities considered are the same as in the example above, and confidence distributions construction are again based on exact and asymptotic pivotal quantities or simulation-based.
 For the Rejection-ABC-type confidence distributions  the tolerance for the discrepancy was still set to $0.1$, the true value of  parameter of interest to $\psi_0=2.6$, and the Huber's tuning constant  to 1.345.
The  proposals for $\psi$ were taken  Uniform in $[-3,9]$, for the parameter $\mu_N$ we sample from a Uniform $[110,130]$, while for $\sigma$ we generated values  from a Uniform $[1,8]$. For the simulations, 4000 values were generated from the proposals and a total of 2000 simulations were performed. 
 
We compare the empirical coverages of $90\%$ and $95\%$ equi-tailed  confidence intervals.  Results are synthetized in Tables \ref{sim_40} and \ref{sim_80}. We also report in Tables  \ref{sim40} and \ref{sim80} the error associated to confidence median point estimators, in terms of bias ($b=\sum_{r=1}^{R}\tilde\theta_r- \theta_0$), probability of overestimation
($PO=\sum_{r=1}^{R}1_{\{\tilde\theta_r> \theta_0\}}$) and I type error with $\alpha=0.05$.
We note that, under the central model, the Wald/Mean CD shows a good performance, as well as some robust CDs (Wald/M-test, CDensity/M-EE and ABC/M-EE). With contaminated data, the Wald/Mean CD tends to be affected by contamination, whereas the robust CDs perform substantially better, with the CDs based on $M$-estimating equations being preferred over those based on $M$-estimators. Asymptotic symmetric confidence densities based on Wald-type robust CDs and ABC-type confidence densities seem to be affected more by bias than the simulated CDs (see Tables \ref{sim40} and \ref{sim80}). Note finally that ABC-type results, even if behaving well, depend on a tolerance choice, hence the results may degradate when the latter is not well calibrated.


As a final remark, note that an interesting aspect of this simulation study was the difference among the approach of using robust $M$-estimating functions instead of robust $M$-estimates, especially in the treatment of nuisance parameters.
 For the M-est CD based on $\tilde\psi$, for each fixed values  $(\psi^*, \lambda^*)$,   the corresponding acceptance probability is  $Pr_{\psi^*, \lambda^*}\left(\tilde\psi^*> \tilde\psi\right) $.  This might be recognized to be similar to a bootstrap $p$-value, with the exception of what is the model considered for the simulation.
 In contrast, for the CD with the profile $M$-estimating function as summary statistic, the acceptance probability for same  values $(\psi^*,\lambda^*)$  is 
$Pr_{\psi^*, \lambda^*} \left(g_\psi(y; \tilde\psi, \lambda^*) \ge 0  \right)$,
as if  $\lambda^*$ was the oracle estimate. Hence, the CD for each $\psi$ is associated to  an average $p$-value, i.e.
$
CD(\psi) =\int Pr_{\psi,\lambda}(g_{\psi}(y; \tilde\psi, \lambda)\geq 0 ) d\lambda.
$
Note that under the true  model, with $(\psi_0, \lambda_0)$ true parameter value, the corresponding $p$-value would be equal to the one without the nuisance parameter
$ Pr_{\psi_0,\lambda_0}(g_\psi(y; \tilde\psi,\lambda_0)\geq 0).$
Note that, instead, keeping fixed the nuisance parameters in the simulations would correspond to consider them as known.
 
\begin{table}[h!]
\centering
\begin{tabular}{|l|l|l|l|l|} \hline
 Contamination &\multicolumn{2}{c|}{0\%} & \multicolumn{2}{c|}{10\%}   \\  
\hline
 $n=40$ & 95\% CI   &  90\%  CI &     95\% CI   &  90\%  CI  \\ \hline 
Wald/Mean& 93.9 &89.1 &   97.1&94.0  \\
Wald/M-test & 93.7 & 88.4 &  94.1&  88.3\\
ABC/Median& 97.1 &93.4 & 97.7  & 93.7 \\
ABC/M-EE & 92.7 &87.2 & 93.7  &88.9  \\
ABC/M-est &97.0 &93.1 & 97.6  & 93.9 \\
CDensity/Median &99.5  & 97.6&  99.2 &98.0  \\
CDensity/M-EE&  95.8&90.5 & 96.7  &  92.1\\
CDensity/M-est &99.4 & 97.3 & 99.2  & 98.0 \\
Boot/basic& 93.4 & 88.0 & 92.3 & 86.1\\
Boot/Norm &93.5 & 88.2 &92.4 & 86.0\\
Boot/Perc & 93.4  & 87.9 &92.3 & 86.1 \\
 \hline
\end{tabular}
\caption{Empirical coverages in a simulation study  without and with
10\% of contamination and $n=40$. }
\label{sim_40}
\end{table}

\begin{table}[h!]
\centering
\begin{tabular}{|l|l|l|l|l|} \hline
Contamination &\multicolumn{2}{c|}{0\%} & \multicolumn{2}{c|}{10\%}   \\  \hline
 $n=80$  & 95\% CI   &  90\%  CI &     95\% CI   &  90\%  CI  \\ \hline 
Wald/Mean&  95.5&90.0& 95.9  &  92.2  \\
Wald/M-test &95.1&89.6& 93.9   &  87.9  \\
ABC/Median&97.2 &  93.5  &   97.3       &93.5     \\
ABC/M-EE &93.3&86.9 &      92.7    &   87.3 \\
ABC/M-est  &97.5  &93.6&    97.2     &   93.9       \\
CDensity/Median & 99.1 &97.6 &  99.2   & 97.5  \\
CDensity/M-EE &95.9&89.4&    96.4   & 92.5    \\
CDensity/M-est& 99.1&97.3&     99.2    & 97.7        \\
Boot/Basic &94.1 &89.5& 92.3& 87.5\\
Boot/Norm & 94.2&89.5& 92.4&87.4\\
Boot/Perc &94.3 &89.6&92.3&87.5\\
 \hline
\end{tabular}
\caption{Empirical coverages in a simulation study  without and with
10\% of contamination and $n=80$. }
\label{sim_80}
\end{table}

\begin{table}[h!]
\centering
\begin{tabular}{ |l|c|c|c|c|c|c|}
  \hline
  Contamination &\multicolumn{3}{c|}{0\%} & \multicolumn{3}{c|}{10\%}   \\  
\hline
  $n=40$ & $|b|$ & $PU$   & I type err.& $|b|$ & $PU$ & I type err.\\ 
  \hline
  Wald/Mean & 0.01  & 0.51  &  0.06  & 5.57   & 0.65 &  0.03\\ 
  Wald/M-test & 0.00  &   0.51 & 0.06 & 0.23  &  0.42& 0.08  \\ 
ABC/Median  & 0.03 &    0.52& 0.01& 0.09  & 0.46 & 0.01 \\ 
  ABC /M-EE  &  0.00 &  0.51  &0.03 & 0.23  & 0.42 & 0.03 \\ 
   ABC/M-est  &  0.01&   0.51 & 0.01& 0.23  & 0.42 & 0.01  \\ 
CDensity/Median  &  0.15&  0.55  &0.01 &  0.03 & 0.51  & 0.01 \\ 
   CDensity/M-EE  &  0.11&   0.55 &0.03 & 0.11  &  0.46& 0.03 \\ 
   CDensity/M-est  &0.13  &0.56    & 0.01& 0.09  &  0.46&0.01  \\ 
Boot/Basic &0.84&0.75&0.06&1.07&0.79&0.10\\
Boot/Norm &0.84&0.75&0.06&1.07&0.79&0.10\\
Boot/Perc&0.84&0.75&0.06&1.07&0.79&0.09\\
   \hline
\end{tabular}
\caption{Measures of stability of CDs: absolute bias ($|b|$), probability of underestimation ($PU$) and I type error $(\alpha=0.05)$ of confidence estimators (medians) in the simulation study with $n=40$.}
\label{sim40}
\end{table}

\begin{table}[h!]
\centering
\begin{tabular}{ |l|c|c|c|c|c|c|}
  \hline
  Contamination &\multicolumn{3}{c|}{0\%} & \multicolumn{3}{c|}{10\%}   \\  
\hline
$ n=80$ & $|b|$ & $PU$   & I type err.& $|b|$ & $PU$ & I type err.\\ 
  \hline
  Wald/Mean   &0.02  & 0.49   &0.05&1.76 &0.58 & 0.05\\ 
  Wald/M-test & 0.01 &0.49&  0.05&0.19 &0.42 &  0.08\\ 
  ABC/Median & 0.02  &  0.50& 0.02   &0.05&0.49&0.02\\ 
   ABC/M-EE &0.02 &  0.48&0.03 & 0.19& 0.42 &0.03\\ 
   ABC/M-est &0.01  &0.49 &0.01&0.19&  0.42& 0.02\\ 
  CDensity/Median & 0.07  &  0.53& 0.01   &0.02& 0.51&0.02\\ 
   CDensity/M-EE &0.08 &  0.53&0.03 & 0.11&0.45 &0.03\\ 
   CDensity/M-est &0.08  &0.54 &0.01 &0.11 & 0.46& 0.02\\ 
Boot/Basic &0.03&51.1&0.05&0.39&0.33&0.09\\
Boot/Norm&0.03&51.1&0.05&0.39&0.33&0.09\\
Boot/Perc&0.03&51.1&0.05&0.39&0.33&0.09\\
   \hline
\end{tabular}
	\caption{Measures of stability of CDs: absolute bias ($|b|$), probability of underestimation ($PU$) and I type error $(\alpha=0.05)$ of confidence estimators (medians) in the simulation study with $ n=80$. }
\label{sim80}
\end{table}

\subsection{Real data application}

\begin{figure}[hb]\center
\includegraphics[scale=0.6]{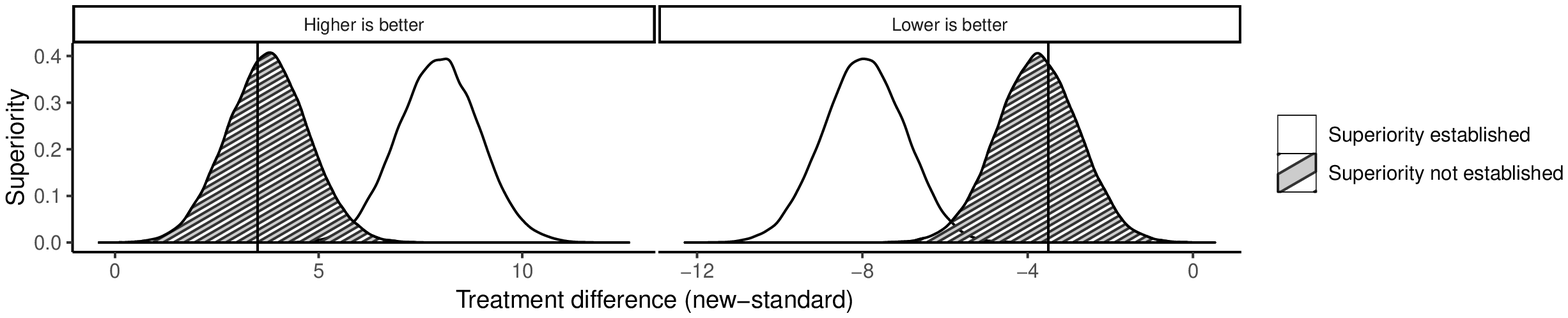}
\caption{Example of making inference with confidence densities in superiority test, with margin $\delta=-3.5$.}
\label{sup}
\end{figure}

A class of problems requiring similar considerations to those of non-inferiority tests, i.e. sensitivity analysis with respect to the reference margin $\delta$, is that of superiority studies (see Fig. \ref{sup}).

Here we analyze the data collected in a randomized controlled    trial  (see Carhart-Harris {\em et al.}, 2021, Nayak {\em et al.},{ 2022})  with the aim of assessing  the superiority of a new therapy with psilocybin (P) versus that with escitalopram (E), in treating major depressive disorder. The dataset contains the scores obtained by $n=57$ patients on a questionnaire, before and after a 6-week period of therapy.
\begin{figure}\center
\includegraphics[scale=0.350]{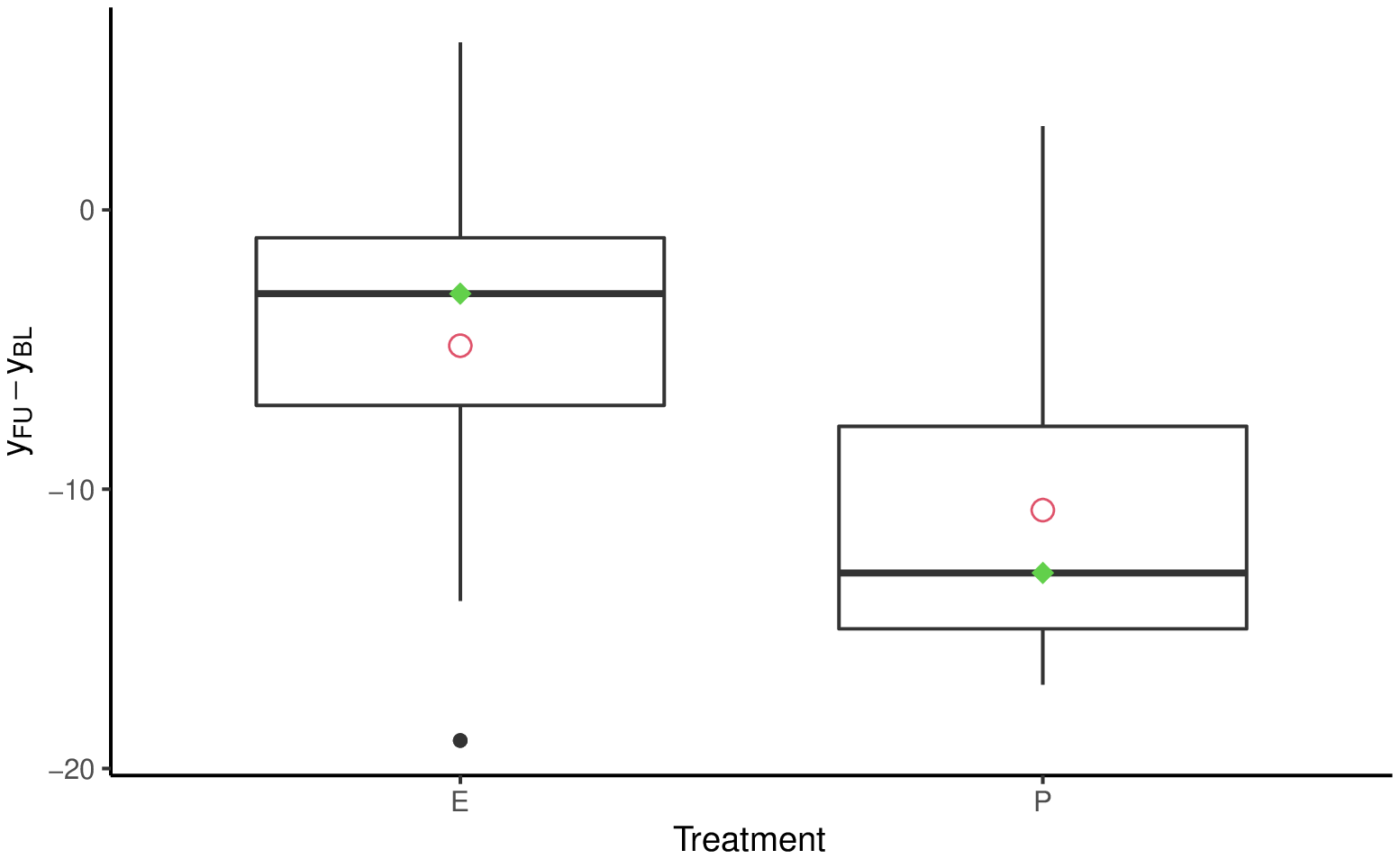}
\includegraphics[scale=0.50]{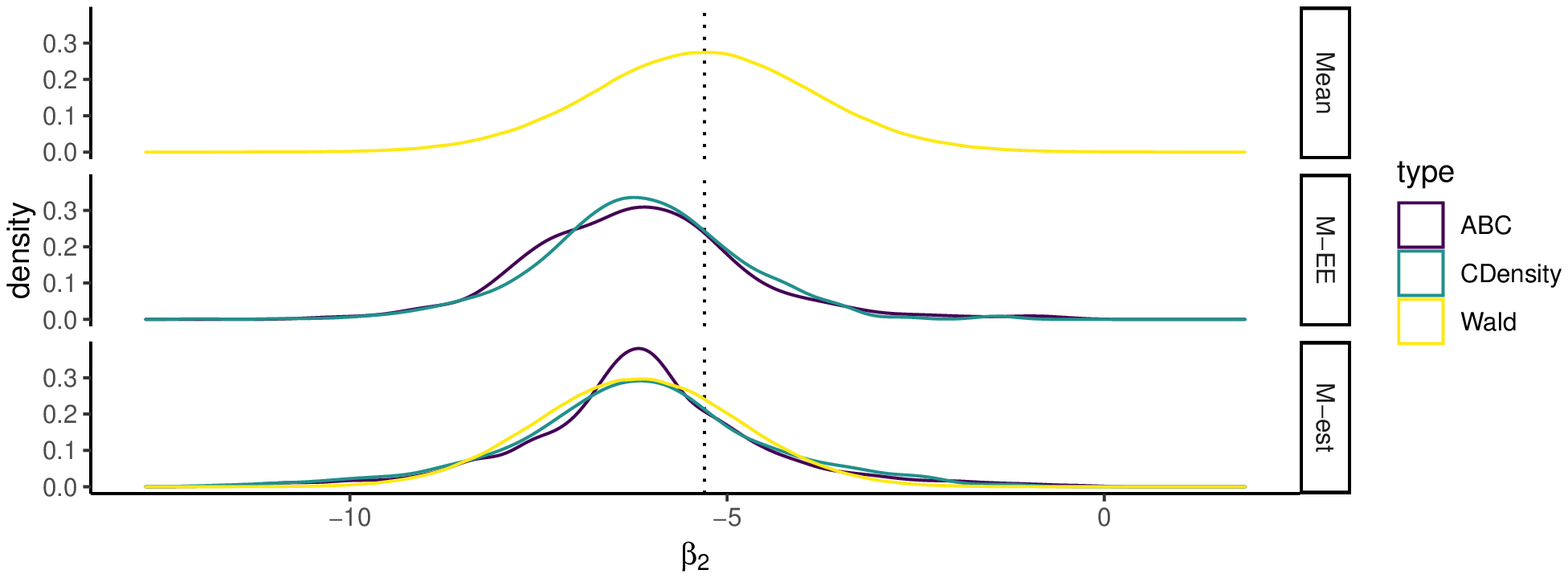}
\caption{Real data example: boxplots of pre-post differences of scores in the   group using  psilocybin (P) versus escitalopram (E) (left),
confidence densitites for the parameter $\beta_2$, measuring the difference in efficacy of the two therapies (right).}
\label{drug_cd}
\end{figure} 
The model considered   for the scores at the time of follow-up (FU)  is the following
$$
y_{FU}=\beta_0+\beta_{1}y_{BL}+\beta_{2}\text{P}+u, \quad u\sim N(0,\sigma^2),
$$
where $y_{BL}$ represents the value at the  baseline and  P is a dummy variable that equals 1 if the subject belongs to the group treated with the new therapy (psilocybin), and thus the coefficient  relates to  the additional change with respect to the control group (escitalopram) after the therapy.  A reduction of the score indicates a clinical improvement; thus superiority is claimed if the estimate of the coefficient $\beta_2$ is sufficiently lower than 0. In particular, in order to conclude in favour of meaningful superiority, the clinicians considered as reference a  margin $\delta=-5.3$.  It is of interest  to provide stable measures of evidence for the statement "$\beta_2>\delta$", with $\delta=-5.3$ ($H_0$). 
 
The MLE for the parameter $\beta_2$ and its standard error are, respectively, $-5.32$ and $1.44$, while the robust  counterparts are $-6.18$ and $1.33$. Note that after removing two outliers the MLE become  -6.23,  with  standard error 1.35. We resume the whole confidence densities based on Wald-type methods together with simulated confidence densities based on Huber's estimators and  Huber's estimating equations in Fig.\ \ref{drug_cd}. As it can be noted the classical confidence density (Wald/Mean) is shifted to the right, because of the presence of outliers. Evidence measures for different margins  are reported in Table \ref{tab_drug}. Using the margin chosen by the clinicians (-5.3) there is no evidence of superiority at  level $\alpha=0.1$; however note that the measure of evidence computed with the Wald-type confidence density (Wald/Mean) is the double of the ones computed with the robust confidence densities.
With a margin of $\delta=-3.5$ all the robust procedure would agree in claiming superiority with $\alpha=0.1$, while according to classical Wald-type confidence density (Wald/Mean) there would not be enough evidence to conclude superiority. 
\begin{table}[ht]
\centering
\begin{tabular}{|l|c|c|c|c|c|}
  \hline
\textbf{$-\delta$} & \textbf{-3.5} & \textbf{-4 }& \textbf{-4.5} & \textbf{-5} & \textbf{-5.3} \\ 
  \hline
 
Wald/Mean & 0.11 & 0.18 & 0.29 & 0.41 & 0.49\\ 
Wald/M-test & 0.02 & 0.05 & 0.10 & 0.19 & 0.26\\ 
ABC/M-est & 0.00 & 0.00 & 0.14 & 0.14 &0.29\\ 
ABC/M-EE & 0.08 & 0.12 & 0.18 & 0.24 &0.30\\ 
CDensity/M-est & 0.03 & 0.09 & 0.14 & 0.21 &0.28\\ 
CDensity/M-EE & 0.07 & 0.14 & 0.17 & 0.22 & 0.28\\ 
    
   \hline
\end{tabular}
\caption{Measures of evidence for the hypothesis "$\beta_2>\delta$"  for several margins. }\label{tab_drug}
\end{table}


\section{Confidence distributions based on integral probability semimetrics}

For mitigating the effects of small departures from model assumptions, and possible dramatic changes of inferential conclusions due to 
 unconvenient choices of pivotal quantities and summaries, the use of non parametric procedures may be an alternative.
 
 Here, we mention the construction of CDs based on  discrepancy measures defined on the space of distributions, belonging to the class of integral probability semimetrics (Muller, 1997) or pseudo-metrics (Huber and Ronchetti, 2009, Chapt.\ 2). These divergences are classically associated to the concept of stability, used as global tests and studied in the context of misspecified models  where the meaningfulness of  model features is uncertain, wherealse directly comparing the distributions happens to be more natural  (Bernton {\em et al.}, 2019, Frazier {\em et al.}, 2020 and Legramanti {\em et al.}, 2022). In particular, we focus on the Kolmogorov-Smirnov distance ($d_{KS}$) and the Wasserstein distance ($d_{W}$) for one-dimensional distributions, defined respectively as 
$$d_{KS}(P,Q)=\underset{y}{\sup} |P(y) - Q(y)|,$$
$$d_{W}(P,Q)=\int_{Y} |P(y) - Q(y)| dy.$$
  The discrepancy  furnishes global indication of potential agreement between the two distributions $P$ and $Q$,  analogously to a likelihood ratio test as in the likelihood-based inference for correctly specified models.
For estimating the distances, empirical cumulative distribution functions  ($\hat P_n(y)$ and $\hat Q_n(y)$, respectively) are used.

In general models, the non-asymptotic distributions of the statistics  $d_{KS}(\cdot, \cdot)$ and $d_{W}(\cdot,\cdot)$ are complex, and numerical methods are employed to compute exact $p$-values.  Here we suggest to rely on Algorithm \ref{algo1}  to derive CDs, once identifyed as  summary statistics suitable discrepancies with distribution stochastically  monotone in a scalar parameter of interest $\theta$. In particular, let us consider  the observed sample $y^{\text{obs}}$  and a fixed reference sample $y^{\text{ref}}$, drawn from a completely known model $f(y;\theta^{\text{ref}})$.
A sequence of unilateral tests can be built by using as observed summary statistic in Algorithm \ref{algo2} the quantity
 $d(y^{\text{obs}},y^{\text{ref}})=d(\hat P(y^{\text{obs}}),\hat Q(y^{\text{ref}}))$, where $d(\cdot,\cdot)$ may be the Kolmogorov-Smirnov distance ($d_{KS}$) or the Wasserstein distance ($d_{W}$).
Then, the CD is obtained with the Accept-Reject scheme of  Algorithm \ref{algo2}, evaluating 
$$
Pr_{\theta^*} (d(y^*,y^{\text{ref}})> d(y^{\text{obs}},y^{\text{ref}})),
$$
where $y^*$ is simulated from the central model  $y^* \sim f(y;\theta^*)$. Also,  by Algorithm \ref{algo3} a confidence density can be retrieved. 
To obtain a proper confidence distribution, the distribution of the summary statistic should be stochastically ordered in the parameter of interest. Hence it is convenient to draw $y^{\text{ref}}$ from the model $f(y;\theta')$,  with    $\theta'$ being the supremum of the proposal distribution  support in Algorithm \ref{algo2}.

Otherwise, a serie of bilateral tests, direclty comparing $d(y^*,y^{\text{obs}})$ to zero,  without a reference sample, can also be performed, for obtaining a confidence curve instead of a proper confidence distribution (Legramanti {\em et al.}, 2022).

\subsection{Example}

As in Legramanti {\em et al.} (2022) we consider a contamination study.
The data  $y^{\text{obs}}=(y_1,\ldots, y_n)$, with sample size  $n=100$, are realizations of a Gaussian random variable $N(\theta,1)$, with nominal value $\theta_0=1$. Within this setting, some scenarios of contamination are investigated: for each one a percentage of observations is substituded with the  most extreme positive realization of a Cauchy of the same size. The amount of contamination here is ${(5\%, 10\%, 15\%)}$, respectively.  In particular, using Algorithm \ref{algo2} we simulated uniformly $\theta$ in $[-3,3]$ and used as a pivot the distance $d(y^{\text {obs}}, y^{\text{ref}} )$, where $y^{\text{ref}}$ is drawn from $N(3,1)$. 

\begin{figure}[h!]\center
\includegraphics[scale=0.7]{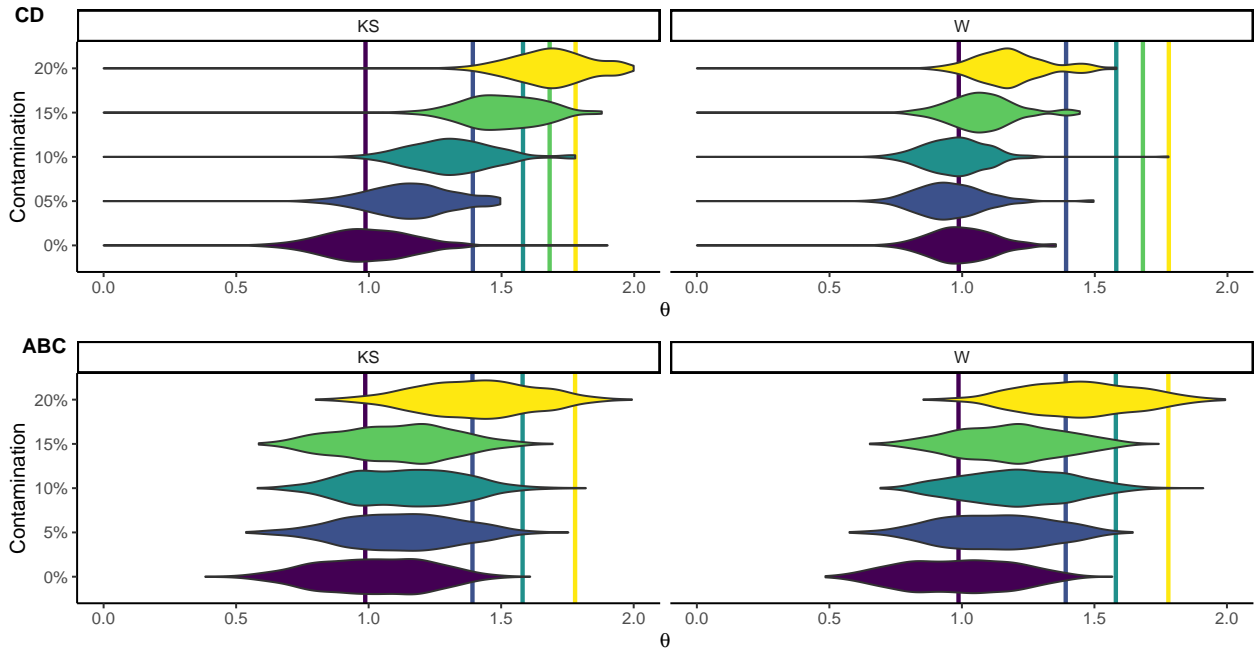}
\caption{Confidence densities and ABC posterior based on Kolmogorv-Smirnov (KS) and Wassserstein (W) distances; vertical lines represent the sample means for increasing level of contamination.}\label{KS-W}
\end{figure}

As shown in Fig.\ \ref{KS-W}, although the sample mean is dragged, the confidence distributions remain close to and concentrated around the nominal value of 1. In particular, the test based on the Wasserstein distance is  higly stable up to the $15\%$ of contamination.
Compared to the approximate posteriors in Legramanti {\em et al.} (2022), the CD based on Wasserstein distance seems even more stable. 
 

Let us denote with  $\tilde\theta^m$ the confidence median and let us focus on the Wasserstein distance. Under  the non contaminated sample ($y_{\theta_0}$) the confidence median satisfies
$$
Pr( d_W(y_{\tilde \theta^m}, y_{\theta^{ref}})>d_W(y_{\theta_0}, y_{\theta^{ref}}))=0.5. 
$$
When considering a $\epsilon$-contaminated sample ($y_{\theta_0^{  c_\epsilon}}$, with $\epsilon<1$\% of the data are not generated from the assumed model), we look for $	\theta^*$ that satisfies
\begin{equation}
Pr( d_W(y_{\theta^*}, y_{\theta^{ref}})>d_W(y_{\theta_0^{c_\epsilon}}, y_{\theta^{ref}}))=0.5. \label{median_cont} 
\end{equation}
The difference $\theta^*-\tilde\theta^m$ is the shift due to the contamination.
Writing  $d_W(y_{\theta_0^{c_\epsilon}},  y_{\theta^{ref}})$ as 
$$  
d_W(y_{\theta_0^{c_\epsilon}},  y_{\theta^{ref}})= (1-\epsilon) d_W(y_{\theta_0}, y_{\theta^{ref}}) + \epsilon \cdot d_W( c,  y_{\theta^{ref}}),  
$$
we can rewrite (\ref{median_cont}) as
$$
Pr \left( d_W(y_{  \theta^*}, y_{\theta^{ref}}) >  d_W(y_{\theta_0}, y_{\theta^{ref}}) +\underbrace{\epsilon[  d_W(c , y_{\theta^{ref}})- d_W(y_{\theta_0}, y_{\theta^{ref}}) ]}_{\Delta} \right)=  0.5.
$$
 As the term $\Delta\rightarrow0$, the confidence median is recovered. In particular this happens in the trivial case, when $\epsilon\rightarrow 0$ or if $\theta^{ref}$   minimizes $ d_W( c , y_{\theta^{ref}})- d_W(y_{\theta_0},y_{\theta^{ref}})$, that means it parametrizes the model which corresponds to the barycenter between the   central one and the model that generates the contamination.
The optimal value cannot be known in advance, but as an initial guess a nonrobust estimate could be considered.

\begin{figure}[h!]\center
 \includegraphics[scale=0.53]{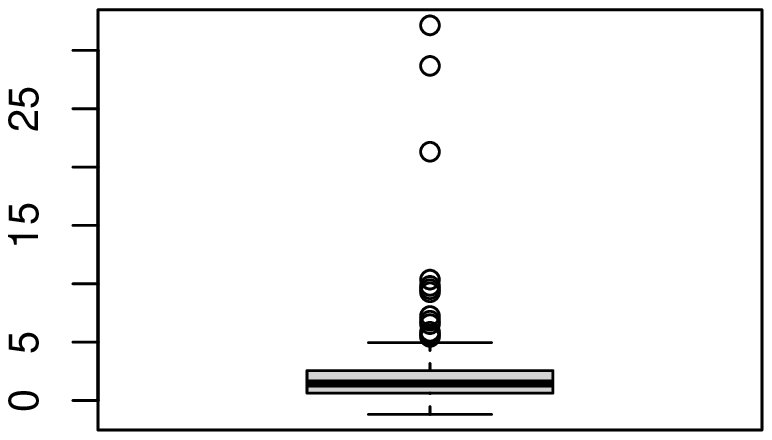}
\includegraphics[scale=0.53]{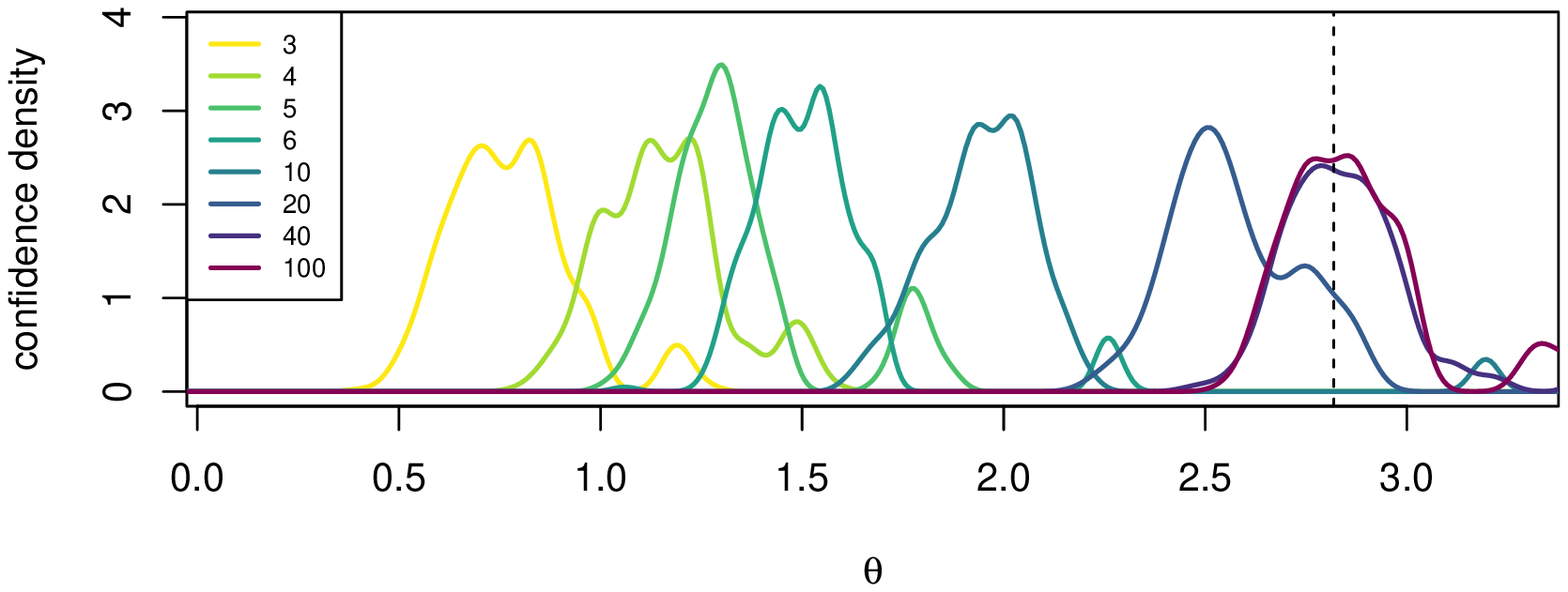}
\caption{Boxplot of the data (left) and confidence densities based on the Wasserstein distance  with $20\%$ of the data ($\epsilon=0.2$) from a contamination model and different choices for the reference parameter. The empirical mean (dotted line) is 2.81, while the true value is 1, and the empirical mean of the uncontaminated sample (1-$\epsilon \; \times 100$ \% of the data) is 1.16.}\label{W_REF}
\end{figure}

For analysing the behaviour of resulting confidence densities under the extreme case in which the contamination amount is $\epsilon=0.2$,  the data  $y^{\text{obs}}$ are still realizations of a Gaussian random variable $N(1,1)$ and for the contamination a percentage of observations is substituded with realizations from a Cauchy. For the derivation of the CDs, we consider different choices for the reference parameter $\theta^{ref}=(3,4,5,6,10,20,40,100)$ (see boxplot of the data and confidence densities in Fig.\ \ref{W_REF}).


\section{Discussion}

Robust statistics is an extension of classical parametric statistics that specifically takes into account the fact that the assumed parametric models used by the researchers are only approximate. The contribution of this paper is to fill the gap between robust inference and confidence distributions analyses. Indeed, in practical applications, CDs are more informative than a simpler confidence interval or a $p$-value, since they describe the complete  distribution estimator for the parameter of interest, as the posterior distribution for Bayesians. In particular, CDs allow to compute measures of evidence for statements of the type "$\psi > \psi_0$" or "$\psi_1 \leq \psi \leq \psi_2$", which are of particular interest in many real data applications, as for instance the one considered in the paper on non-inferiority testing. Remark that the application to non-inferiority trials discussed here can be easily extended to the comparison of other parameters than means, such as odds ratios, hazard ratios, etc., where stability of inferential conclusions with respect to small changes in the data or to small model departures is essential.

The derivation of robust CDs discussed in Section 3.2, based on a frequentist reinterpretation of ABC techniques to obtain a normalized pseudo-likelihood function for the parameter of interest, represents a practical alternative to other robust pseudo-likelihood functions, such as the empirical likelihood or the quasi-likelihood (see e.g. Greco {\em et al.}, 2008, and references therein). Indeed, the aforementioned functions may present some drawbacks: the empirical likelihood is not computable for small sample sizes and quasi-likelihoods can be easily obtained only for scalar parameters. On the contrary, the approach that directly incorporates robust estimating functions into ABC techniques, with respect to available approaches based on pseudo-likelihoods, can be computationally faster when the evaluation of the estimating function is expensive, can be computed even for small sample sizes and for multidimensional parameters of interest, and the derived normalized pseudo-likelihood has the right curvature (see Ruli {\em et al.}, 2020). Obviously, the proposed method can be applied to any unbiased robust estimating equations, such as $S$-estimating equations.

Related to the above comment, we highlight that for a scalar parameter of interest in the presence of nuisance parameters we propose to modify the algorithm of Ruli {\em et al.} (2020) by using the a profile estimating equation and plugging in the value of proposals for nuisance parameters used to generate pseudo-data. The treatment of nuisance parameters is a theme of ever-renewed attention in frequentist inference. An interesting extension of our proposal, not only limited to robust inference but for general estimating functions, is the further study of ways to handling nuisance parameters with CD-based inference, in particular with the options available with the bootstrap {(see DiCiccio and Romano, 1988)} and with ABC techniques based on general profile-type estimating functions.

As a final remark, we note that in this paper we mention the possibility to adopt  non parametric criteria and statistics, other than just centrality measures,  for deriving confidence distributions for a scalar parameter of interest in presence of contamination.
A central parametric model is assumed but  the observed data are evaluated in terms of  non parametric pseudo-distrances from a reference model, directly based on the empirical cdfs.
At present our preliminary study is limited to   models with a scalar parameter of interest, since 
adapting these kind of test procedures to more complex models to deal with nuisance parameters may require information about the context, situation-dependent considerations and also give rise to confidence curves  instead of confidence distributions.


\section*{References}

\begin{description}
\item
Barndorff-Nielsen, O.E., Cox, D.R. (1994). {\em Inference and Asymptotics}, CRC Press.
\item
Bee, M., Benedetti, R., Espa, G. (2017). Approximate maximum likelihood estimation of the Bingham distribution. {\em Computational Statistics \& Data Analysis}, {\bf  108}, 84--96.
\item 
Bortolato, E., Ventura, L. (2022). Confidence distributions and fusion inference for intractable likelihoods. {\em Book of Short Papers SIS 2022}, Caserta 2022, to appear.
\item
Brazzale, A.R., Davison, A.C., Reid, N. (2007). {\em  Applied asymptotics. Case-studies in small sample statistics}. Cambridge University Press.
\item Bernton, E., Jacob, P. E., Gerber, M.,  Robert, C. P. (2019). On parameter estimation with the Wasserstein distance. {\em Information and Inference: A Journal of the IMA}, {\bf 8}, 657--676.
\item
Carhart-Harris, R., Giribaldi, B., Watts, R., Baker-Jones, M., Murphy-Beiner, A., Murphy, R.,  Nutt, D. J. (2021). Trial of psilocybin versus escitalopram for depression. {\em New England Journal of Medicine}, {\bf 384}, 1402--1411.
\item
Chen, X., Zhou,  W. (2020).  { Robust inference via multiplier bootstrap}. {\em The Annals of Statistics}, {\bf 48}, 1665--1691.
\item
D'Agostino, R.Bb, Massaro, J.M., Sullivan, L.M. (2003). Non- inferiority trials: design concepts and issues the encounters of academic consultants in statistics. {\em Statistics in Medicine}, {\bf 22}, 169--186.
\item
Dawid, A.P., Musio, M., Ventura, L. (2016). Minimum scoring rule inference. {\em Scand.\ J.\ 
Statist.}, {\bf 43}, 123--138.
 \item
DiCiccio, T. J.,  Romano, J. P. (1988). A review of bootstrap confidence intervals. {\em Journal of the Royal Statistical Society: Series B (Methodological)}, {\bf 50}, 338--354.
\item
DiCiccio, T. J., Efron, B. (1996). Bootstrap confidence intervals. {\em Statistical science}, {\bf 11}, 189-228.
\item
Efron, B. (1979). Bootstrap methods: another look at the jackknife. { \em The Annals of Statistics}, {\bf 7}, 1--26.
\item
Farcomeni, A., Ventura, L. (2012). An overview of robust methods in medical research. {\em Statistical Methods in Medical Research}, {\bf 21}, 111--133.
\item
Field, C.A., Ronchetti, E. (1991). {\em Small Sample Asymptotics}. IMS Monograph Series, Hayward (CA).
\item
Frazier, D. T., Robert, C. P.,  Rousseau, J. (2020). Model misspecification in approximate Bayesian computation: consequences and diagnostics.{ \em Journal of the Royal Statistical Society: Series B (Statistical Methodology)}, {\bf 82}, 421--444.
\item
Garret, A.D. (2003). Therapeutic equivalence: fallacies and falsification. {\em Statistics in Medicine},  {\bf 22}, 741--762.
\item
Ghosh, M., Basu, A. (2013). Robust estimation for independent non-homogeneous observations using density power divergence with applications to linear regression. {\em Electronic Journal of Statististics}, {\bf 7}, 2420--2456.
\item
Greco, L., Racugno, W., Ventura L. (2008). Robust likelihood functions in Bayesian inference. {\em Journal of Statistical Planning and Inference}, {\bf 138}, 1258--1270.
\item
Hampel, F.R., Ronchetti, E.M., Rousseeuw, P.J., Stahel, W.A. (1986). {\em Robust Statistics. The Approach Based on Influence Functions}. Wiley.
\item
Heritier, S., Cantoni, E., Copt, S., Victoria-Feser, M.P. (2009). {\em Robust Methods in Biostatistics}. Wiley.
\item
Heritier, S. and Ronchetti, E.M.(1994). Robust bounded-influence tests in general parametric models.  {\em Journal of the American Statististical Association}, {\bf 89}, 897--904.
\item
Hjort, N.L., Schweder, T. (2018). Confidence distributions and related themes. {\em Journal of Statististical Planning and Inference}, {\bf 195}, 1--13.
\item
Huber, P.J., Ronchetti, E.M. (2009). {\em Robust Statistics}. Wiley, New York.
\item 
Legramanti, S., Durante, D., Alquier, P. (2022). Concentration and robustness of discrepancy-based ABC via Rademacher complexity, {\em arXiv:2206.06991}.
\item
Lyddon, S. P., Holmes, C. C.,  Walker, S. G. (2019). General Bayesian updating and the loss-likelihood bootstrap. {\em Biometrika}, {\bf 106}, 465--478.
\item
Muller, A. (1997). Integral probability metrics and their generating classes of functions. {\em Advances in Applied Probability}, {\bf 29}, 429--443.
\item
Nayak, S., Bari, B. A., Yaden, D. B., Spriggs, M. J., Rosas, F., Peill, J. M., ... , Carhart-Harris, R. (2022). A Bayesian Reanalysis of a Trial of Psilocybin versus Escitalopram for Depression. {\em PsyArXiv preprint}
\item
Newton, M. A.,  Raftery, A. E. (1994). Approximate Bayesian inference with the weighted likelihood bootstrap. {\em Journal of the Royal Statistical Society: Series B (Methodological)}, {\bf 56}, 3--26.
\item
Reid, N. (2003). The 2000 Wald memorial lectures: asymptotics and the theory of inference. {\em Annals of Statistics}, {\bf 31}, 1695--1731.
\item
Ronchetti, E., Ventura, L. (2001). Between stability and higher-order asymptotics. {\em Statistics and Computing}, {\bf 11}, 67--73.
\item
Rothmann, M.D., Wiens, B.L., Chan, I.S.F (2012). {\em Design and Analysis of Non-Inferiority Trials}. CRC Press.
\item
Rubio, F.J., Johansen, A.M. (2013). A simple approach to maximum intractable likelihood
estimation. {\em Electronic Journal of Statistics}, {\bf 7}, 1632--54.
\item
Ruli, E., Sartori, N., Ventura, L. (2016)- Approximate Bayesian Computation with composite score functions. {\em Statistics and Computing}, {\bf 26}, 679--692.
\item
Ruli, E., Sartori, N., Ventura, L. (2020). Robust approximate Bayesian inference. {\em Journal of Statistical Planning and Inference}, {\bf 205}, 10--22.
\item
Ruli, E., Ventura, L. (2021). Can Bayesian, confidence distribution and frequentist inference agree?. {\em Statistical Methods \& Applications}, {\bf 30}, 359--373.
\item
Ruli, E., Ventura, L., Musio., M. (2022). Robust confidence distributions from proper scoring rules. {\em Statistics}, {\bf 56}, 455-478.
\item
Schweder, T., Hjort, N.L. (2016). {\em Confidence, Likelihood, Probability: Statistical Inference with Confidence Distributions}. Cambridge University Press.
\item
Severini, T.A. (2000). {\em Likelihood methods in statistics}. Oxford University Press.
\item
Soubeyrand, S., Haon-Lasportes, E., (2015). Weak convergence of posteriors conditional on maximum pseudo-likelihood estimates and implications in
ABC. {\em Statistics and Probability Letters}, {\bf 107}, 84--92.
\item
Thornton, S., Li, W., Xie, M. (2022), Approximate confidence distribution computing, {\em 	arXiv:2206.01707}.
\item
Tsallis, C. (1988). Possible generalization of Boltzmann-Gibbs statistics. {\em Journal of Statistical Physics}, {\bf 52}, 479--487.
\item
Varin, C., Reid, N., Firth, D. (2011). An overview of composite likelihood methods. {\em Statistica Sinica}, {\bf 21}, 5--42.
\item
Ventura, L., Racugno, W. (2016). Pseudo-likelihoods for Bayesian inference. In: {\em Topics on Methodological and Applied Statistical Inference, Series Studies in Theoretical and Applied Statistics}, Springer-Verlag, 205--220.
\item
Ventura, L., Sartori, N., Racugno, W. (2013). Objective Bayesian higher-order asymptotics in models with nuisance parameters. {\em Computational Statistics and Data Analysis}, {\bf 60}, 90--96.
\item
Xie, M., Singh, K. (2013). Confidence distribution, the frequentist distribution estimator of a parameter: a review. {\em International Statistical Review}, {\bf 81}, 3--39.
\end{description}

\end{document}